\documentclass[pdflatex,sn-nature]{sn-jnl}

\usepackage{graphicx}%
\usepackage{multirow}%
\usepackage{amsmath,amssymb,amsfonts}%
\usepackage{amsthm}%
\usepackage{mathrsfs}%
\usepackage[title]{appendix}%
\usepackage{xcolor}%
\usepackage{textcomp}%
\usepackage{manyfoot}%
\usepackage{booktabs}%
\usepackage{algorithm}%
\usepackage{algorithmicx}%
\usepackage{algpseudocode}%
\usepackage{listings}%
\usepackage{mathrsfs}

\theoremstyle{thmstyleone}%
\theoremstyle{thmstyletwo}%

\theoremstyle{thmstylethree}%

\begin{document}


\title[Article Title]{Leveraging neural network interatomic potentials for a foundation model of chemistry}


\author[1]{\fnm{So Yeon} \sur{Kim}} 

\author*[1]{\fnm{Yang Jeong} \sur{Park}}\email{parkyj@mit.edu}

\author*[1,2]{\fnm{Ju} \sur{Li}}\email{liju@mit.edu}

\affil[1]{\orgdiv{Department of Nuclear Science and Engineering}, \orgname{Massachusetts Institute of Technology}, \orgaddress{\street{77 Massachusetts avenue}, \city{Cambridge}, \postcode{02139}, \state{MA}, \country{United States}}}


\affil[2]{\orgdiv{Department of Materials Science and Engineering}, \orgname{Massachusetts Institute of Technology}, \orgaddress{\street{77 Massachusetts avenue}, \city{Cambridge}, \postcode{02139}, \state{MA}, \country{United States}}}



\abstract{Large-scale foundation models, including neural network interatomic potentials (NIPs) in computational materials science, have demonstrated significant potential. However, despite their success in accelerating atomistic simulations, NIPs face challenges in directly predicting electronic properties and often require coupling to higher-scale models or extensive simulations for macroscopic properties. Machine learning (ML) offers alternatives for structure-to-property mapping but faces trade-offs: feature-based methods often lack generalizability, while deep neural networks require significant data and computational power. To address these trade-offs, we introduce HackNIP, a two-stage pipeline that leverages pretrained NIPs. This method first extracts fixed-length feature vectors (embeddings) from NIP foundation models and then uses these embeddings to train shallow ML models for downstream structure-to-property predictions. This study investigates whether such a hybridization approach, by ``hacking" the NIP, can outperform end-to-end deep neural networks, determines the dataset size at which this transfer learning approach surpasses direct fine-tuning of the NIP, and identifies which NIP embedding depths yield the most informative features. HackNIP is benchmarked on Matbench, evaluated for data efficiency, and tested on diverse tasks including \textit{ab initio}, experimental, and molecular properties. We also analyze how embedding depth impacts performance. This work demonstrates a hybridization strategy to overcome ML trade-offs in materials science, aiming to democratize high-performance predictive modeling.}  

\keywords{Materials design, Structure-to-property prediction, Neural network interatomic potentials, Feature-based machine learning}



\maketitle

\section{Introduction}\label{sec1}  
Large-scale foundation models developed through self-supervised learning have demonstrated great potential across numerous applications, ranging from text generation and language comprehension to image generation and interpretation~\cite{radford2018GPT1, radford2019GPT2, radford2021CLIP, touvron2023Llama, touvron2023llama2, jablonka2024LeveragingLLMforChem}. These models follow ``a scaling law," where performance improves continuously with increased training compute. In computational materials science, a prominent example is machine learning interatomic potentials (MLIPs), which learn to approximate high-fidelity potential energy surfaces from quantum-mechanical data, serving as fast, accurate surrogates for expensive first-principles calculations. Particularly, MLIP foundation models based on deep neural networks (NNs)---known as neural network interatomic potentials (NIPs)---like GNoME~\cite{merchant2023GNoME} and MatterSim~\cite{yang2024MatterSim} have confirmed the scaling law, and being trained on large-scale density functional theory (DFT) datasets such as MPtrj~\cite{deng2023CHGNet} and OMat24~\cite{barroso-luque2024OpenMaterials2024}, these NIPs can replicate DFT results at high accuracy at just a fraction of computational costs previously required, assuming we treat the costs of constructing DFT datasets as sunk. 

The performance gains in large-scale NIP foundation models \cite{batatia2022MACE, merchant2023GNoME, liao2023EquiformerV2,deng2023CHGNet, yang2024MatterSim, neumann2024ORBv2, rhodes2025ORBv3} have expanded the spatiotemporal limits of atomistic simulations achievable at DFT‐level accuracy. Nevertheless, many practical challenges remain in downstream structure‐to‐property predictions. First, because NIPs bypass explicit computation of electron density and directly map to energies, they often struggle to predict electronic properties. Second, predicting macroscopic properties still tend to require coupling to higher‐scale models (e.g., coarse‐grained potentials) or generating enormous MD ensembles. Third, bridging the gap between DFT predictions and experimental measurements for properties that are not directly accessible via atomistic simulations remains a key development goal.

\begin{figure*}[!htp]
    \centering
    \includegraphics[width=0.8\textwidth]{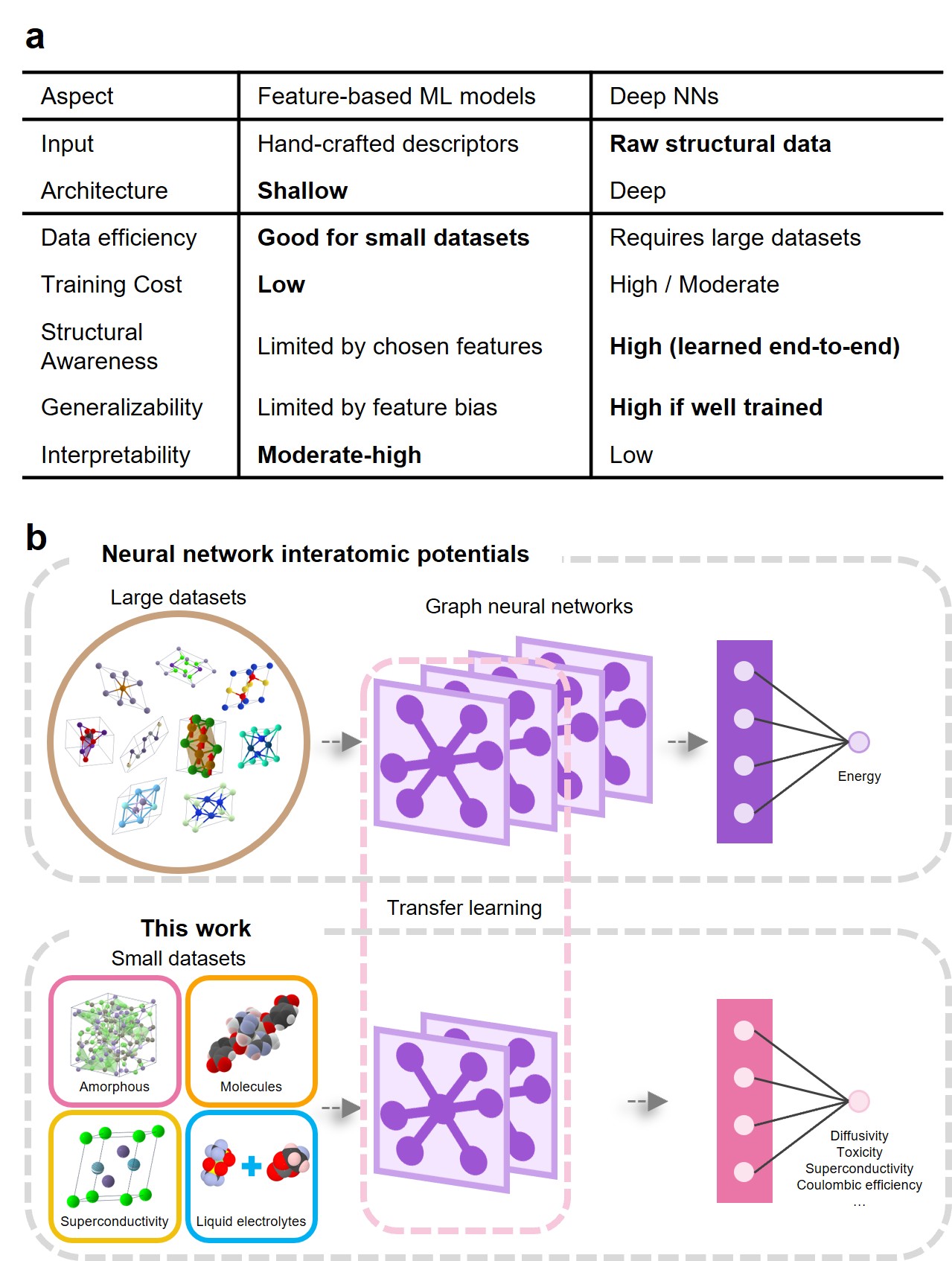}
    \caption{\textbf{Overview of machine learning strategies for structure-to-property predictions in materials science.} \textbf{a,} Comparison of conventional feature-based machine learning (ML) models and deep neural networks (NNs). The comparative advantages of each model relative to the other and the architectural characteristics that provide these advantages are highlighted in bold. \textbf{b,} Transfer learning using pretrained neural network interatomic potentials (NIPs) in this work. Traditional NIPs (top) are trained on large datasets comprising diverse crystalline structures to learn interatomic interactions via deep NN. This work (bottom) applies transfer learning to small datasets spanning various material classes—including amorphous solids, molecular systems, superconductors, and liquid electrolytes—by leveraging pretrained representations that enable accurate modeling in data-scarce regimes.}
    \label{fig:hackNIP}
\end{figure*}

Machine learning (ML) presents a viable alternative for high-throughput mapping of structure to properties in cases where atomistic simulations with NIPs encounter limitations. Figure \ref{fig:hackNIP}a summarizes the architectural characteristics and trade-offs between conventional feature-based ML and deep NNs. Conventional feature-based ML models use fixed, hand-crafted descriptors and shallow architectures, making them efficient on small datasets and interpretable (though still far less so than atomistic simulations), but limited in generalizability due to the expense of structural detail and inherent feature biases. In contrast, deep NNs---like graph neural networks (GNNs)---can learn directly from atomic structures, capturing complex patterns with broad generalizability, but they demand more data and computational resources and are often difficult to interpret.

One promising strategy involves extracting features from pretrained universal NIPs and feeding them into lightweight ML models for downstream tasks~\cite{shiota2024UniversalNeuralNetworkasDescriptors}. This hybridization approach can potentially overcome those trade-offs, given the fact that they can use raw structural data as input and employ shallow architectures for property mapping (Figure \ref{fig:hackNIP}a). Such an approach raises three key questions: First, can a shallow ML model with NIP-derived embeddings outperform end-to-end deep NNs for property prediction, by harnessing broader knowledge of the base NIP---namely ``hacking" the NIP? Second, at what dataset size range does this transfer-learning pipeline surpass direct fine-tuning of the same NIP? Third, which embedding depths of an NIP yield the most informative features for downstream tasks?

To explore these questions, we introduce HackNIP, a two-stage pipeline (Figure \ref{fig:hackNIP}b): (1) feature extraction from NIP foundation models to convert atomic structures into fixed-length vectors; and (2) structure-to-property prediction using shallow ML models trained on the extracted feature vectors paired with target properties. First, we benchmark its performance on the standard Matbench~\cite{dunn2020MatBench} dataset. Second, we assess data efficiency by evaluating model performance across varying dataset sizes, and compare it to that of a direct fine-tuning approach. Third, we apply the method to diverse downstream tasks---encompassing dynamic properties from \textit{ab initio} calculations, experimental datasets involving one or more materials, and molecular properties---to demonstrate generalizability. Finally, we discuss how the choice of embedding depth for feature extraction, together with NIP characteristics, influence performance, providing guidance for future development. Overall, this work demonstrate the potential of hybridization approaches to overcome trade-offs in ML for materials science and to democratize high-performing ML capabilities.

\section{Results}
\subsection{Benchmarking model performance on Matbench tasks} 
We first benchmarked HackNIP on the Matbench test suite, an automated leaderboard for ML algorithms predicting a diverse range of solid‐state properties. Matbench comprises 13 tasks over 10 datasets, varying in target property, sample size, and data-generation method. From these, we selected all regression tasks with available atomic‐structure inputs. The target properties of these tasks are: exfoliation energy ($E_\mathrm{exfoliation}$); frequency of the last phonon peak in the phonon density of states ($\nu$); refractive index ($n$); logarithmic shear and bulk moduli ($G$ and $K$, respectively); perovskite formation energy (E$_\mathrm{f, perovskite cell}$); band gap ($E_\mathrm{g}$); and general formation energy ($E_\mathrm{f}$). Sample counts ranged from 636 (exfoliation energy) to 132,752 (formation energy). All structures and target values in the datasets were compiled by DFT calculations. We used the five predefined train/test splits provided by the Matbench Python package. The optimized hyperparameters for each task are summarized in Supplementary Table \ref{tab:s_matbench_opted-hp}. Various combinations of NIP foundation models (ORB, Equiformer, and MACE) and shallow regressors (MODNet, XGBoost, and MLP) were evaluated (Figure \ref{fig:NIP-ml_combination}), and the best-performing pair---ORB–MODNet---was selected for demonstration.

\begin{figure*}[!htp]
    \centering
    \includegraphics[width=\textwidth]{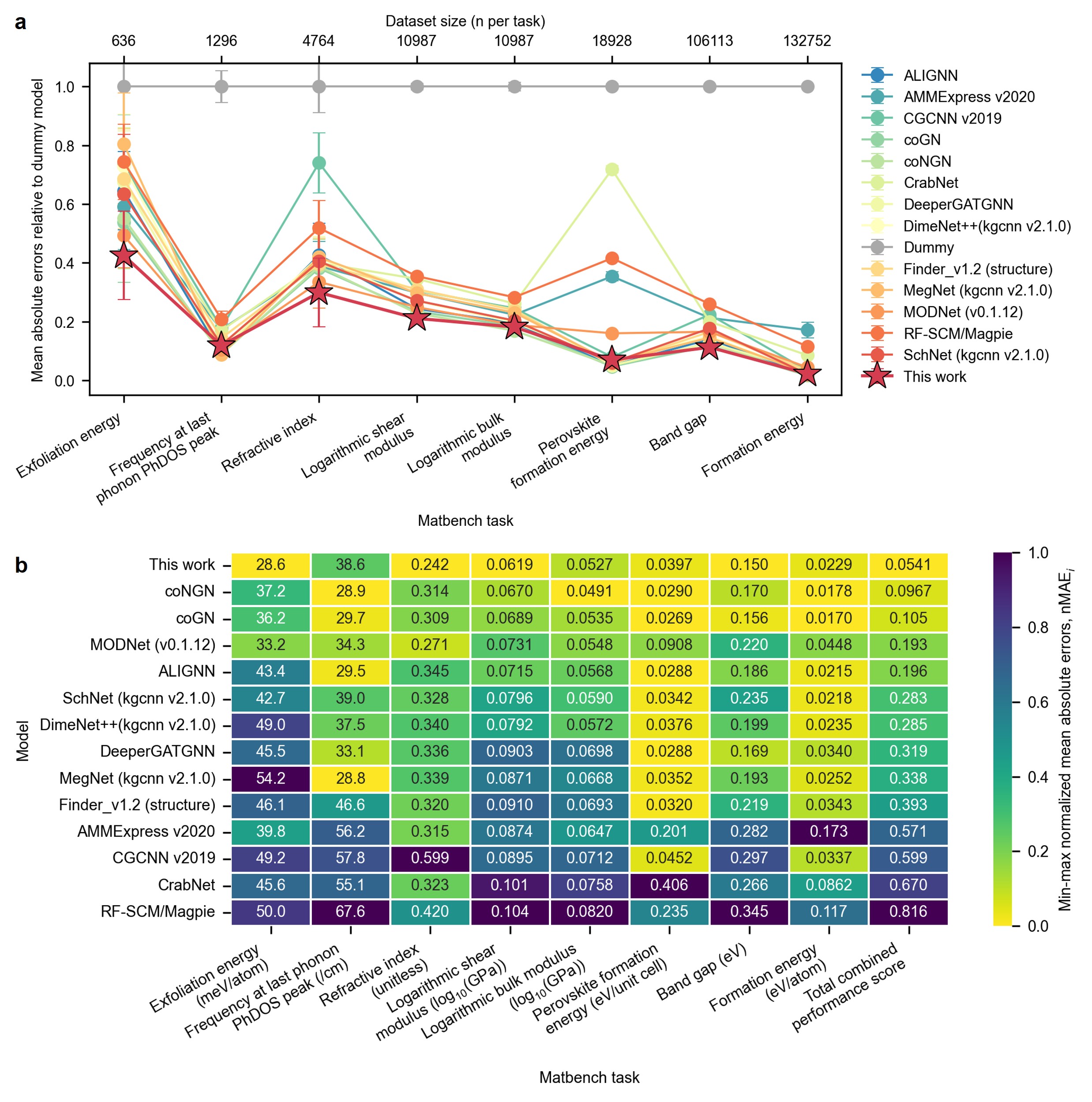}
    \caption{\textbf{Benchmarking model performance across Matbench tasks.} \textbf{a,} Mean absolute error relative to a dummy model that predicts the mean of the dataset. Dataset size for each task is indicated above the plot. The performance of HackNIP (red stars) is compared to those of baseline and state-of-the-art models. Error bars represent standard deviations over five-fold experiments. Lower mean absolute error indicates better performance. \textbf{b,} Raw mean absolute error of each model. Cells are color-coded by the scaled MAE, ranging from 0 (best) to 1 (worst), calculated using min–max normalization within each task.}
    \label{fig:matbench_benchmarking}
\end{figure*}

Figure \ref{fig:matbench_benchmarking}a shows the relative mean absolute errors (MAEs) of HackNIP and selected baselines compared to a dummy model, which always predicts the training‐set mean. The Matbench leaderboard (as of May 2025) comprises 28 community‐contributed models, each evaluated on a subset of tasks; here, we retained only those reporting results across all eight regression tasks. These baselines include graph neural networks (e.g., CGCNN and ALIGNN) as well as feature‐based ML models (e.g., AMMExpress and MODNet). We denote the MAE of model $i$ on fold $j$ as $\mathrm{MAE}_i^j$, and its average across folds as $\mathrm{MAE}_i^{\mathrm{avg}}=\frac{1}{5}\sum_{j=1}^5\mathrm{MAE}_i^j$. The relative MAE is then defined as $\mathrm{MAE}_i^{\mathrm{avg}}$ divided by $\mathrm{MAE}_\mathrm{dummy}^{\mathrm{avg}}$ where $\mathrm{MAE}_\mathrm{dummy}^{\mathrm{avg}}$ is the average MAE of the dummy model on five folds. HackNIP achieves the state-of-the-art predictive performance with the lowest nMAE on four tasks and remains competitive elsewhere, delineating an approximate Pareto-optimal front across the eight Matbench tasks with varying dataset sizes.

Figure \ref{fig:matbench_benchmarking}b shows the absolute MAEs, with each cell color-coded according to the mix-man normalized MAE of each model $i$ (nMAE$_i$) defined as 

\begin{equation}
 \mathrm{nMAE}_{i}
  = \frac{\mathrm{MAE}_i^{\mathrm{avg}} - \min\limits_k \mathrm{MAE}_k^{\mathrm{avg}}}
                          {\max\limits_k \mathrm{MAE}_k^{\mathrm{avg}} - \min\limits_k \mathrm{MAE}_k^{\mathrm{avg}}},   
\end{equation}

\noindent where the extrema run over all non‐dummy models, $k$. Cells are colored by $\mathrm{nMAE}_{i}$ between 0 (best) and 1 (worst). HackNIP attains $\mathrm{nMAE}_{i}=0$ for four tasks and remains below 0.11 for three, indicating it remains competitive with end-to-end deep NNs. Although HackNIP underperforms some recent GNNs on the last‐phonon‐peak frequency and perovskite formation energy tasks, it still outperforms all feature‐based ML models and earlier GNN architectures. Notably, it achieves a high coefficient of determination $R^2=0.968\pm0.010$ for the phonon-peak frequency (see parity plot in Figure \ref{fig:s_matbench_parity}) and an MAE of 40 meV/unit cell for perovskite formation energy---below the so-called chemical accuracy (1 kcal/mol or 43 meV/atom)---indicating it predicts DFT reference values more accurately than DFT itself predicts experimental formation energies. Conversely, for refractive index, HackNIP attains the lowest MAE but a poor $R^2$ ($0.334\pm0.270$), likely due to data sparsity at high index values; logarithmic scaling might mitigate this, but no additional preprocessing was applied to preserve comparability. Averaging $\mathrm{nMAE}_{i}$ over the eight tasks yields a combined performance score, for which HackNIP ranks first among all leaderboard models as shown in Figure \ref{fig:matbench_benchmarking}a.

\subsection{Data efficiency}  

To examine data efficiency, we conducted HackNIP training and ORB fine-tuning on the five Matbench tasks with more than 10$^4$ samples---shear and bulk moduli, perovskite formation energy, band gap, and general formation energy. For property-specific fine-tuning, we attached a single-output MLP head (see Methods for more details) and unfroze all ORB encoder layers except the initial edge encoder and the final decoder to preserve interatomic embeddings and read-out mappings that generalize across chemistries, thereby minimizing disturbance to rich, physics-informed representations learned during construction of the NIP foundation model. This targeted freezing mitigates catastrophic forgetting, stabilizes gradients during training, and concentrates learning capacity on adapting only the most relevant internal layers and our lightweight head.

\begin{figure*}[h]
    \centering
    \includegraphics[width=\textwidth]{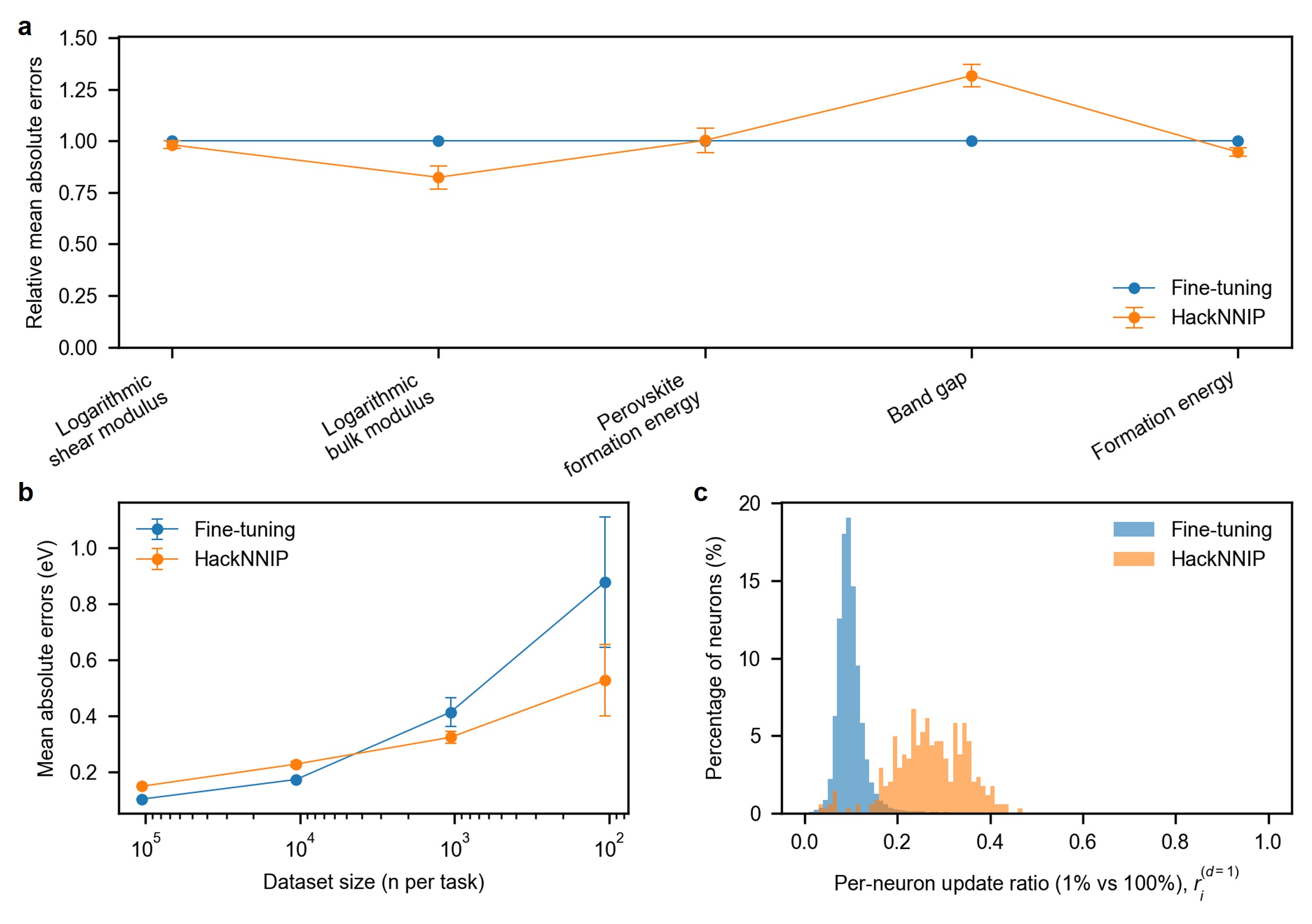}
    \caption{\textbf{Comparison of fine-tuning approach and HackNIP across varying dataset sizes.} \textbf{a,} Mean absolute errors of HackNIP relative to fine-tuned NIP across five Matbench tasks when using $\sim$10$^4$-sample subsets: shesar and bulk moduli (10,987 samples), perovskite and general formation energies (10,000 samples), and band gap with (10,611 samples). \textbf{b,} Raw mean absolute errors of fine-tuning and HackNIP as a function of dataset fraction (0.1\%, 1\%, 10\%, and 100\%). Error bars represent standard deviations over five independent random subsets (for fine-tuning) or five five-fold experiments (for HackNIP). \textbf{c,} Probability distribution of neurons plotted against per-neuron update ratio, comparing models trained on a 1\% subset to those trained on the full (100\%) dataset for each approach. Blue: fine-tuned ORB; Orange: HackNIP pipeline consisting of ORB and MODNet.}
    \label{fig:matbench_scaling}
\end{figure*}

Figure \ref{fig:matbench_scaling}a compares the relative MAE of HackNIP against fine-tuned ORB when each is trained on approximately $\sim$10$^4$ samples---10,987 samples for moduli (full dataset); 10,611 for band gap (10\% subset); 10,000 for formation energies (53\% of the perovskite formation-energy dataset; 7.5\% of the general formation-energy dataset). For each subset, the desired fraction of the full dataset was drawn by random sampling. The ORB embedding depths used for feature extraction are identical to those used in Figure \ref{fig:matbench_benchmarking}, although MODNet hyperparameters were re-optimized for the smaller subsets. The performance of HackNIP matches or exceeds that of fine-tuned ORB in four of the five tasks---band gap being the sole exception---suggesting that it can serve as a competitive universal predictor up to dataset sizes of $\sim$10$^4$ without costly end-to-end fine-tuning for each property. 

Interestingly, while HackNIP and fine-tuned ORB performed similarly when predicting shear modulus and formation energies, they differed substantially in predicting bulk modulus and band gap. Specifically, HackNIP outperformed direct fine-tuning by 18\% on bulk modulus, even though this was the task where HackNIP exhibited the poorest performance among the leaderboard models showing lowest nMAE across the five tasks with $>$10$^4$ datapoints (Figure \ref{fig:matbench_benchmarking}b). In contrast, HackNIP underperformed compared to fine-tuned ORB for band gap despite HackNIP topping the leaderboard (Figure \ref{fig:matbench_benchmarking}b). These differences may be attributed to specific characteristics of the ORB architecture.

We then further examined the band gap task to identify the dataset size at which HackNIP begins to outperform fine-tuned ORB. Figure \ref{fig:matbench_scaling}b plots MAE versus the total number of samples used for training and testing, with the x-axis shown on a logarithmic scale. Performance of two approaches was compared at four dataset sizes: 0.1\%, 1\%, 10\%, and 100\% of the full dataset. At 0.1\% ($\sim$10$^2$ samples) and 1\% ($\sim$10$^3$ samples) of the dataset, we performed fine-tuning on five different random sampled subsets to ensure statistical robustness. With 100\% of the data, fine-tuned ORB reaches an MAE of $\sim$0.10 eV, surpassing all Matbench leaderboard entries, including Microsoft’s fine-tuned MatterSim (MAE = 0.12 eV). As the training fraction decreases, the MAE of fine-tuned ORB grows roughly as a power law, whereas HackNIP’s MAE increases approximately log-linearly. Consequently, the two curves cross between 10$^3$ and 10$^4$ samples, with HackNIP increasingly outperforming fine-tuned ORB as the dataset size decreases. 


Lastly, to investigate how much the weights move under limited‐data versus full‐data training, we analyzed the update ratio. One can compare models of identical architecture trained on different dataset sizes at various granularities (per‐weight, per‐neuron, per‐layer, etc.). Here, we use the per‐neuron update ratio, which offers a balanced granularity and interpretability, capturing meaningful weight shifts at the neuron level without the noise inherent in per-weight comparisons or the coarseness of per-layer measures. Specifically, if the weight tensor for a given layer is $\mathbf{W}: \mathbb{R}^{\mathrm{in}} \;\longrightarrow\; \mathbb{R}^{\mathrm{out}}$, then its $i$-th row, denoted $\mathbf{W}_i$, contains all of the incoming weights to neuron $i$. The per-neuron update ratio is then defined by

\begin{equation}
    \Delta_i^{(d)}
    = \mathbf{W}_i^{(d)} - \mathbf{W}_i^{(\mathrm{pre})},
    \quad d \in \{1\%\,,\,10\%\,,\,100\%\}, \\[6pt]
\end{equation}

\begin{equation}
    r_i^{(d)}
    = \frac{\|\Delta_i^{(d)}\|_2}
           {\|\Delta_i^{(100\%)}\|_2 + \varepsilon}
    \;=\;
    \frac{\|\ \mathbf{W}_i^{(d)} - \mathbf{W}_i^{(\mathrm{pre})}\|_2}
           {\|\ \mathbf{W}_i^{(100\%)} - \mathbf{W}_i^{(\mathrm{pre})}\|_2 + \varepsilon},   
\end{equation}

\noindent where $\Delta_i^{(d)}$ is the change in neuron $i$’s incoming weight vector when training on fraction $d$ of the data, with $\mathbf{W}_i^{(\mathrm{pre})}$ denoting either the pretrained baseline weights (for ORB fine-tuning approach) and or randomly initialized baseline weights of MODNet (for HackNIP approach). The $r_i^{(d)}$ is the normalized update ratio, measuring how much of the full‐data update each neuron achieves on fraction $d$, with a small $\varepsilon$ to avoid division by zero. The closer $r_i^{(d)}$ is to 1, the more neuron $i$ moves when trained on $d\%$ of the data compared to the full dataset, indicating that it is data‐efficient. In contrast, $r_i^{(d)} \ll 1$ suggests that neuron $i$ barely changes under limited data, so it requires more examples before it ``wakes up.''

Figure \ref{fig:matbench_scaling}c shows the probability distribution of neurons as a function of the per-neuron update ratio, computed with the model trained on 1\% of the data in the numerator ($d = 1$) and 100\% in the denominator. This lets us quantitatively compare the data efficiency of weight updates. The HackNIP curve is shifted to the right relative to fine-tuning, indicating that the 1\% HackNIP model more closely matches its 100\% counterpart than the 1\% fine-tuned model does its 100\% counterpart. Specifically, while only half of the neurons in the fine-tuned model update beyond 0.1, almost all neurons in HackNIP undergo updates greater than 0.1, with the median update ratio being 0.3. This greater inertia is expected given the much larger network size (42,393 neurons for ORB fine-tuning versus just 343 for HackNIP) and suggests that fine-tuning necessitates freezing a larger proportion of parameters when data is scarce, increasing the need for human intervention and domain knowledge.

\subsection{Generalizability}  

All target properties in the aforementioned Matbench benchmarking tasks are zero-temperature DFT properties. In contrast, real‐world materials research often requires ML models for finite‐temperature, dynamic properties---whose datasets are far more time‐consuming to assemble even with high‐throughput workflows---as well as those for properties that cannot be computed directly from first principles or even those for system‐level performance metrics arising from interactions among multiple materials. To demonstrate the broad applicability of HackNIP, we evaluated it on a diverse collection of literature datasets: (1) Li diffusivity in amorphous materials (regression), computed via \textit{ab initio} molecular dynamics (AIMD) \cite{zheng2024abinitioDiffusivity}; (2) superconducting critical temperature (regression), measured experimentally \cite{sommer20233DSCDatasetSuperconductors}; (3) Coulombic efficiency of Li‐ion batteries (regression), involving combinations of electrolyte precursors \cite{kim2023ElectrolyteDesign}; and (4--6) molecular-property classification tasks such as beta-secretase 1 inhibition (BACE) \cite{subramanian2026BACEdataset}, blood–brain barrier penetration (BBBP) \cite{martins2012BBBPdataset}, and clinical toxicity (ClinTox) \cite{gayvert2016ClinToxdataset} from the MoleculeNet database \cite{wu2018MoleculeNet}.

\begin{figure*}[h]
    \centering
    \includegraphics[width=\textwidth]{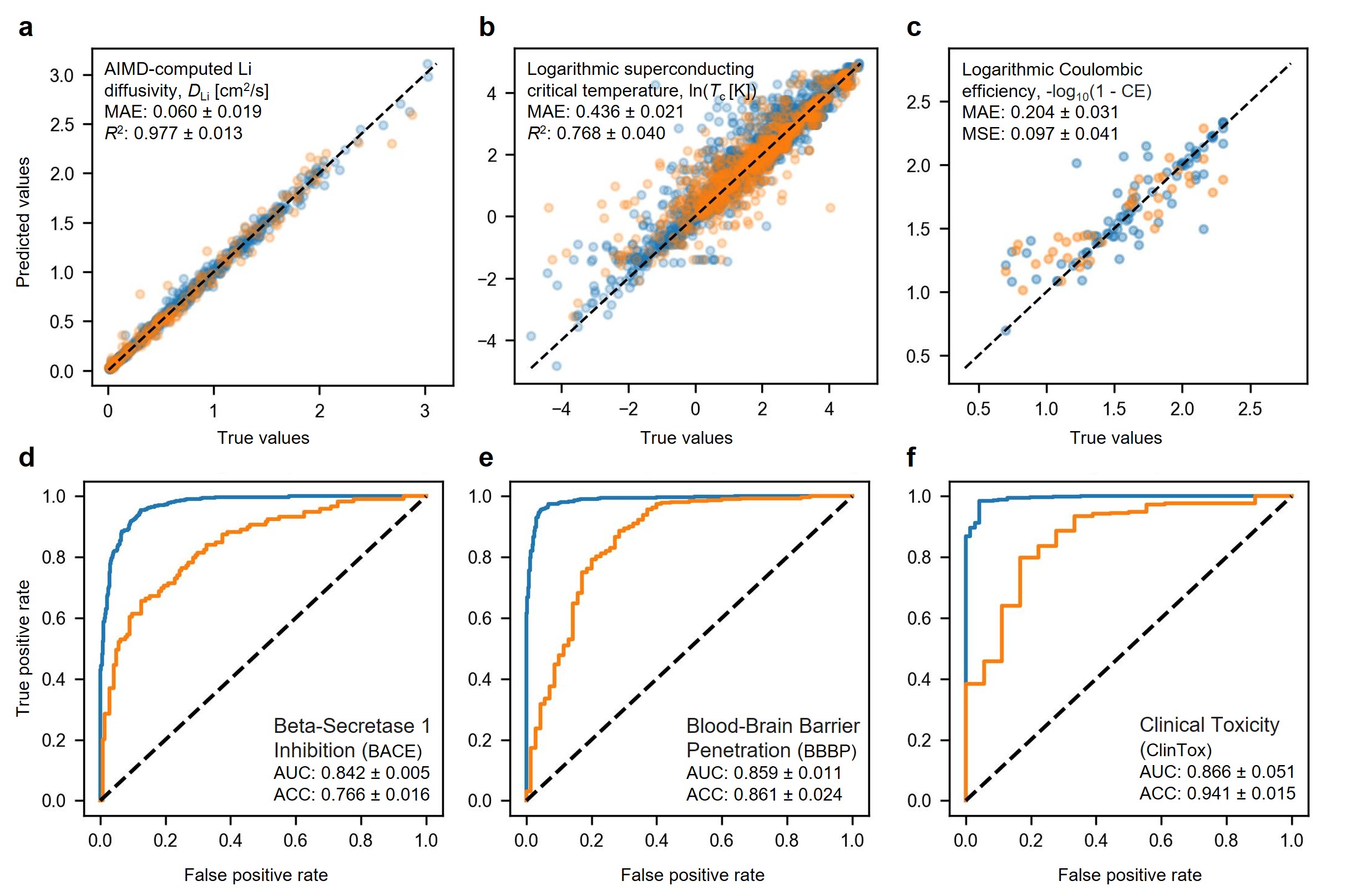}
    \caption{\textbf{Model performance across diverse downstream tasks.} Parity plots for regression tasks on: \textbf{a,} Li diffusivity calculated by \textit{ab initio} molecular dynamics simulation; \textbf{b,} superconducting critical temperatures; \textbf{c,} Coulombic efficiency. Mean absolute errors (MAEs) and either the coefficients of determination ($R^2$) or mean squared errors (MSEs)---matching the metrics used in previous studies---are noted. Receiver operating characteristic curves for classification tasks on: \textbf{d,} beta-secretase 1 inhibition; \textbf{e,} blood-brain-barrier penetration (BBBP); \textbf{f,} clinical toxicity (ClinTox). Area under the curve (AUC) and accuracy (ACC) values are noted. Blue: train; Orange: test; Dashed black line: identity line.}
    \label{fig:downstream_tasks}
\end{figure*}

Figure \ref{fig:downstream_tasks} presents parity plots for all regression tasks and receiver operating characteristic (ROC) curves for the classification tasks. For each regression dataset, we report MAE alongside either $R^2$ or mean squared errors (MSEs), matching the metrics used in previous studies in the literature. All three regression targets were trained on logarithmically scaled outputs, while plots and metric calculations use whichever scale (linear, natural log, or base-10 log) the relevant literature employed, to facilitate direct comparison. For the Li diffusivity and superconducting critical temperature tasks, we performed five‐fold experiments using an 80/20 train/test split created with a Matbench seed of 18012019. For the Coulombic efficiency task, we conducted five independent evaluations using 70/30 train/test splits with five randomly chosen seeds (due to its limited dataset size and to match prior studies). The optimized hyperparameters are summarized in Table \ref{tab:s_general_opted-hp}. We discuss the predictive performance for each task below. 


Figure \ref{fig:downstream_tasks}a displays a parity plot of the model performance on AIMD-computed Li diffusivity. The database includes but not limited to Li-containing amorphous solids, and the diffusivities of constituent elements were computed at high temperatures ranging from 1000 K to 5000 K. Among these, we used every Li-containing data point in the database \cite{zheng2024abinitioDiffusivity} and included temperature as an additional feature alongside ORB embeddings. For direct comparison with the literature---which likely excluded the 5000 K data, perhaps because many compounds are no longer in the condensed phase at that temperature (though this is not stated explicitly), we drew the parity plot on a linear scale and estimated predictive performance in this representation, yielding MAE = 0.060 $R^2$ = 0.977. In comparison, Ref. \cite{zheng2024abinitioDiffusivity} used human-engineered descriptors and reported MAE = 0.081 $R^2$ = 0.956 with a random forest regressor and MAE = 0.083 $R^2$ = 0.960 with a XGBoost regressor, both using 85/15 train/test splits. It is notable that HackNIP can achieve state-of-the-art performance for finite-temperature, dynamic properties as well without hand-engineered features---which might require substantial prior knowledge---and thus without human bias.

Figure \ref{fig:downstream_tasks}b shows HackNIP's performance on predicting superconducting critical temperatures ($T_\mathrm{c}$), which must be determined experimentally rather than computed from first principles. We used the 3DSC database \cite{sommer20233DSCDatasetSuperconductors}, which augments the SuperCon composition entries with approximate three-dimensional crystal structures drawn from the Materials Project, wherever possible. This merged dataset comprises more than 5,000 entries. After applying a base-10 logarithm to the critical-temperature values, we removed any entries with non-finite targets, yielding a final training set of approximately 4,000 entries. The parity plot is drawn in natural log scale following Ref. \cite{stanev2018machine}. In the low-$T_\mathrm{c}$ regime, deviations from the parity line were relatively large, resulting in MAE = 0.436 and $R^2$ = 0.768, while at higher levels the predictions align closely with the identity line, which is the region of greatest practical interest. We note that Ref. \cite{stanev2018machine} reported a higher $R^2$ = 0.88 with composition-based features from Matminer \cite{ward2018matminer} and a random-forest regressor, while direct comparison to such ML studies is challenging, since most employ larger SuperCon subsets (12,000+) via composition-only featurization, adopt different target ranges ($ln (T_\mathrm{c})$: 2.3–5 vs. –4–4 here), and employ different train/test splits (85/15 in Ref. \cite{stanev2018machine} vs. our 80/20). 

Figure \ref{fig:downstream_tasks}c illustrates the results for Coulombic efficiency of Li-ion batteries with organic liquid electrolytes. We used the experimental dataset \cite{kim2023ElectrolyteDesign}, which contains 124 entries. Since each electrolyte formulation can involve multiple precursor species---solvents, solutes, additives, etc.---mixed in different proportions, we first convert them to molar concentrations. Then, the embedding of each compound was computed, and a weighted sum of these embeddings was calculated using the molar concentrations as coefficients to obtain the final embedding. HackNIP achieved superior predictive performance, with an MAE of 0.204 and an MSE of 0.097---about a 30\% reduction compared to the minimum MSE of 0.333 attained by linear, random-forest, boosting, or bagging models using human-engineered features.

For classification tasks on molecular data, we plot true-positive rate versus false-positive rate (ROC curves)---a threshold-agnostic metric that is robust to class imbalance---and report area under curve (AUC) values in addition to accuracy (ACC) values. Figure \ref{fig:downstream_tasks}d--f shows the classification performance for BACE, BBBP, and ClinTox, respectively. Those MoleculeNet benchmark datasets are highly relevant to real-world drug discovery. For example, BACE1 is a key enzyme involved in producing $\beta$‐amyloid peptides, which aggregate in the brains of Alzheimer’s patients. Inhibiting BACE1 has long been considered a promising strategy for slowing Alzheimer’s progression. Whether a small molecule can penetrate the blood–brain barrier (BBB) is critical for developing central nervous system drugs (e.g., for Parkinson’s, Alzheimer’s, depression). The ClinTox dataset collects clinical trial outcomes (approved vs. withdrawn/toxic) for thousands of drugs and drug‐like molecules. 

Since the molecular data are provided as Simplified Molecular Input Line Entry System (SMILES) strings, we first generated three-dimensional structures and performed geometry optimization before feature extraction. As with regression, we extracted features from ORB, performed layer‐search with fixed MODNet hyperparameters, then optimized those hyperparameters and evaluated via five-fold experiments. In all three cases, AUC values are in the mid-80\% range, indicating strong classification performance. This result is comparable to pretrained models such as MolCLR \cite{wang2022MolCLR} and GEM \cite{fang2022GEM}, which have been specifically trained to extract useful representations for molecules, whereas ORB was trained to predict energy using a supervised-learning approach.

\subsection{NIP embedding depth for feature extraction and model performance}  

Next, we examine the effect of NIP embedding depths used for feature extraction on HackNIP performance. The ORB (orb-v2 in this work) backbone comprises 15 layers, each of which can be used to extract embeddings. Obtaining embeddings from the last layer requires propagating raw coordinates and species labels through the shared encoder and all 15 backbone layers, which substantially increases featurization cost—especially when processing on the order of $>$10$^4$ samples. As summarized in Table \ref{tab:s_matbench_opted-hp}, most Matbench regression tasks attain their best MAE using early-layer embeddings (1 $\leq L \leq$ 7), with exfoliation energy as the sole exception (optimal at $L = 11$). Likewise, Li diffusivity, superconducting critical temperature, and Coulombic efficiency reach minimal MAE at $L = 2,4,1$, respectively (Table \ref{tab:s_general_opted-hp}). By contrast, molecular‐property classification tasks (BACE, BBBP, ClinTox) achieve peak AUC at deeper layers ($L = 10,15,13$), as shown in Table \ref{tab:s_general_opted-hp}.

\begin{figure*}[htp!]
    \centering
    \includegraphics[width=\textwidth]{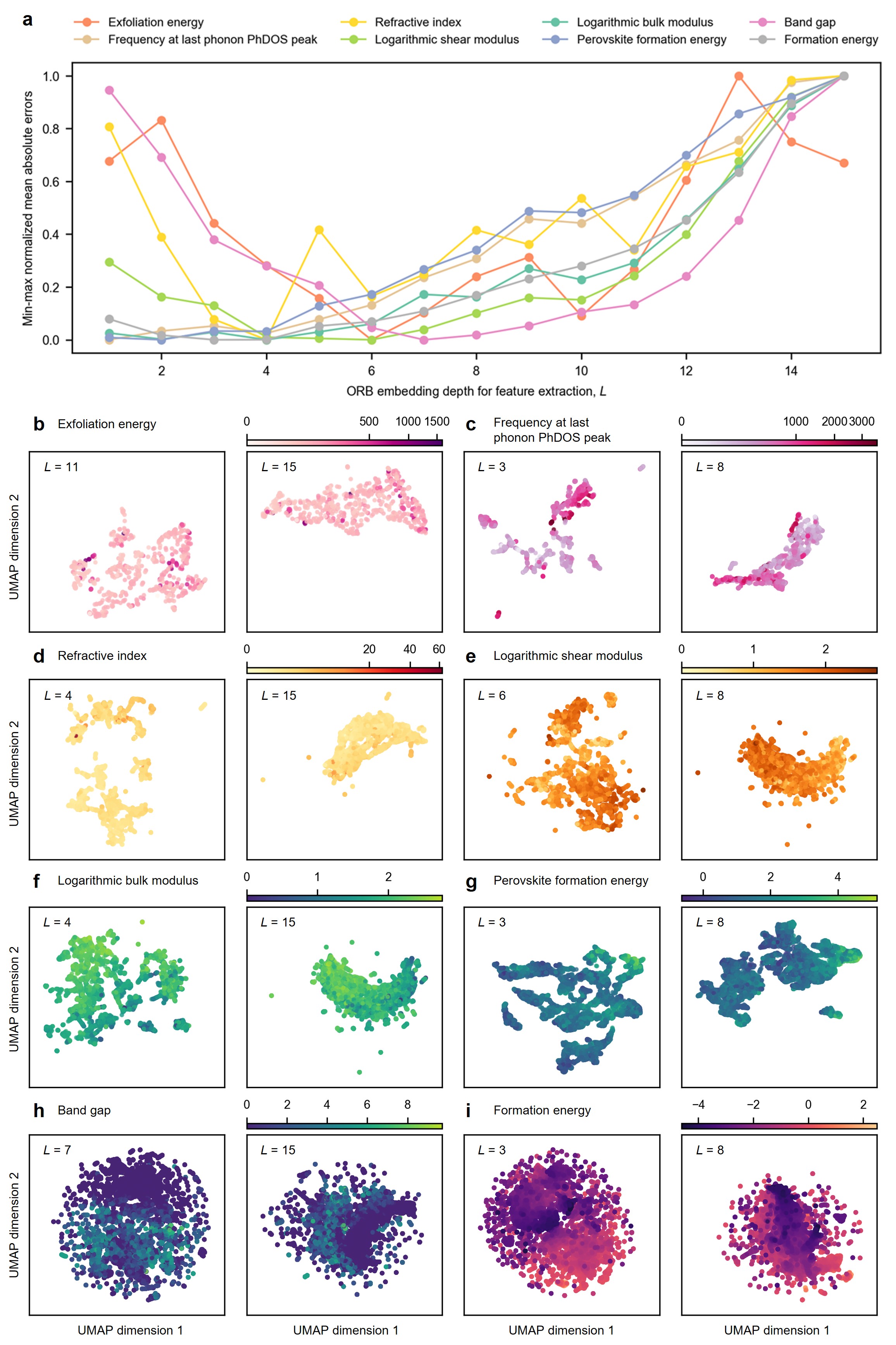}
    \caption{\textbf{Embedding depth used feature extraction and model performance.} \textbf{a,} HackNIP performance as a function of ORB embedding depth used for feature extraction. Mean absolute errors are min-max normalized across embedding depths for each Matbench task. Five-fold experiments were performed for each point and only the mean values are presented for clarity. \textbf{b--i,} Uniform manifold approximation and projection (UMAP) of ORB embeddings for each Matbench task. ORB embeddings from the best‐performing layer ($L^*$) and the final layer ($L=15$) are reduced from 256 dimensions to two dimensions for each task, with color indicating the target value. Depending on the target‐value distribution, the color scheme is either power‐normalized or linearized.
}
    \label{fig:matbench_layer}
\end{figure*}

First, we investigated how HackNIP’s MAE varies with the ORB embedding depth ($L$) used for feature extraction on Matbench tasks (Figure \ref{fig:matbench_layer}a), employing MODNet with a fixed hyperparameter without further optimization. To compare across tasks, we scaled each MAE between its task-specific minimum and maximum, defining the mix-max normalized MAE for task $t$ using embeddings from layer $L$ (nMAE$_{t,L}$) as

\begin{equation}
  \mathrm{nMAE}_{t,L}
  = \frac{\mathrm{MAE}_{t,L} - \min\limits_{L'} \mathrm{MAE}_{t,L'}}
         {\max\limits_{L'} \mathrm{MAE}_{t,L'} - \min\limits_{L'} \mathrm{MAE}_{t,L'}}
  \,,
  \label{eq:xxx_scaled_mae}
\end{equation}

\noindent where $\mathrm{MAE}_{t,L}$ is the average MAE for task $t$ at layer $L$ over five-fold experiments. The minimum and the maximum are taken over all layers $L'$ for the same task. In most cases, embeddings from the final layer yield the highest (i.e., worst) MAE, whereas optimal performance is achieved by using embeddings from the first half of the ORB network for feature extraction. The optimal layer shifts depending on the downstream ML algorithm as shown in Figure \ref{fig:s_layer_xgb-mlp} for two additional HackNIP combinations (ORB + XGBoost and ORB + MLP); however, both combinations still exhibit increasing MAE for later layers.

To visualize how the layer-wise embeddings evolve through the NIP network, we projected them into two dimensions (2D) using Uniform Manifold Approximation and Projection (UMAP). Figures \ref{fig:matbench_layer}b--i show the 2D UMAPs of embeddings from the best performing layer $L^*$ for each Matbench task and the last layer ($L=15$). In most cases, the final-layer embeddings collapse into a highly condensed manifold, indicating that the shared encoder and deep layers have compressed the representation into a low-dimensional subspace. In contrast, embeddings extracted from the layers that yielded the best predictive performance exhibit a much more diffuse UMAP layout; clusters corresponding to different regions of the property space are more widely separated, and outliers stand out more clearly. When starting from these less-compressed embeddings, a downstream model (i.e., MODNet) might better harness material characteristics that were de-emphasized (and thus suppressed) during construction of NIP foundation model, thereby leading to improved performance when using shallower layers of a given NIP architecture for feature extraction.


Lastly, to investigate possible reasons for wider variation of best-performing layers observed for diverse downstream tasks, we examined the Jensen-Shannon Divergence. The Jensen-Shannon Divergence (JSD) is a method used in probability theory to measure the similarity between two probability distributions. The JSD for two distributions $P$ and $Q$ is calculated as the average of the Kullback-Leibler (KL) divergence of each distribution to a mixture distribution $M = \frac{1}{2}(P+Q)$:

\begin{equation}
  JSD(P||Q) = \frac{1}{2} D_\mathrm{KL}(P||M) + \frac{1}{2} D_\mathrm{KL}(Q||M),  
\end{equation}

\noindent where $D_\mathrm{KL}$ denotes the KL divergence. The JSD is always non-negative, and its value is 0 if and only if $P$ and $Q$ are identical. It also has the advantages of being symmetric.

\begin{figure*}[h!]
    \centering
    \includegraphics[width=\textwidth]{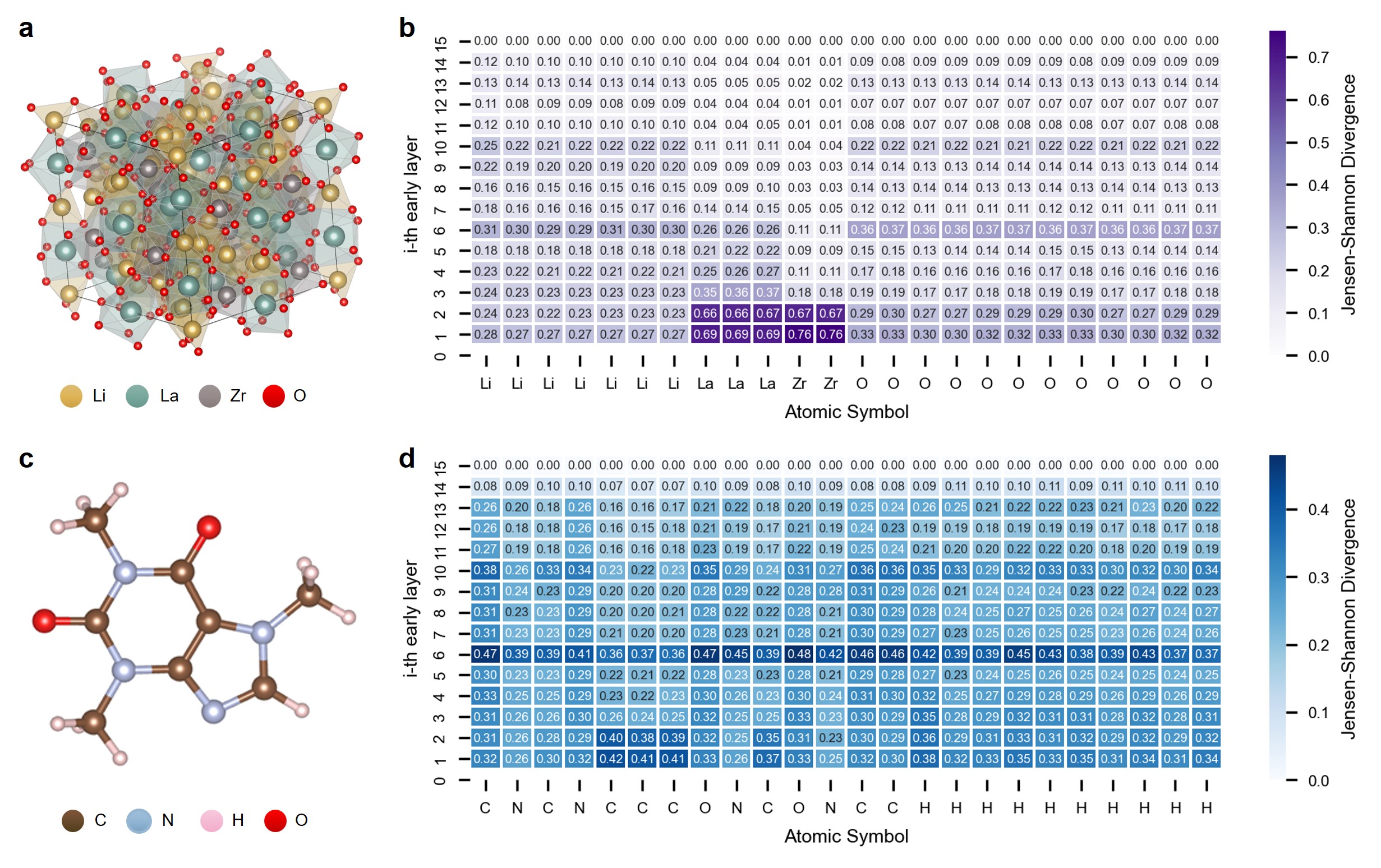}
    \caption{\textbf{Jensen–Shannon divergence (JSD) distributions across embedding depths} \textbf{a,} Atomic structure of the solid electrolyte, Li$_7$La$_3$Zr$_2$O$_{12}$ (LLZO). \textbf{b,} Per-atom JSD values vs. embedding depth for LLZO, illustrating that early transition metals exhibit very high divergence in layers 1--3 before rapidly diminishing. \textbf{c,} Molecular structure of caffeine, C$_8$H$_{10}$N$_4$O$_2$. \textbf{d,} Per-atom JSD values vs. embedding depth for caffeine, showing that divergence remains relatively high through deep layers. This contrast correlates with the finding that the best-performing layer for Li diffusivity prediction is shallow, whereas optimal layers for molecular-property prediction in caffeine lie much deeper.}
    \label{fig:layer_jsd}
\end{figure*}

Figures \ref{fig:layer_jsd}a,b display the atomic structure of the representative solid electrolyte, Li$_7$La$_3$Zr$_2$O$_{12}$ (LLZO). Figures \ref{fig:layer_jsd}c,d show the molecular structure of caffeine, C$_8$H$_{10}$N$_4$O$_2$, and the corresponding JSD distribution across embedding depths for each atom constituting the molecule. For both examples, the average JSD values across all atoms are high at embedding depths of 6 and 10. Meanwhile, in the LLZO example, JSD values for early transition metals---which shape the Li-conduction pathways---are very high in the first two to three layers, quickly diminishing afterward. This suggests that the conclusion for the final prediction is already made very early in the neural network. This behavior may be correlated with the finding that the best-performing layer for Li diffusivity prediction is $L = 2$, which is relatively shallow. On the other hand, in the caffeine example, the JSD values remain relatively high into deeper layers (e.g., $L = 13$), compared to those of LLZO. This suggests that more message passing and global information are needed to make a final conclusion. This behavior may be related to the observation that the best-performing layers for molecular properties are relatively deep. An in-depth analysis of the JSD distribution, combined with feature-importance analysis available for shallow models, may offer deeper insights into structure–property relationships—for instance, by identifying the most critical atoms and their JSD distribution across embedding depths.

\section{Discussion}\label{sec12} 

Materials design is often a multi-objective optimization problem, where one must improve certain properties without degrading others \cite{olson1997computational}. Historically, designers have been driven to establish interdisciplinary knowledge \cite{fine1994materials} and handle a larger number of elements to strike the right balance among multiple properties \cite{oh2019engineering}, as exemplified by ``high-entropy" materials (alloys, ceramics, catalysts, etc.) \cite{han2024multifunctional}. That strategy creates an effectively infinite design space, yet available data remain scarce because both computational and experimental investigations of these materials are often resource-intensive. Accordingly, ML models that are both universal and data-efficient are critical for transforming materials design practices, and lightweight yet robust approaches are essential to democratize state-of-the-art methods for academic/industrial groups with relatively limited computational resources. As demonstrated by its benchmarking performance and scaling capability, HackNIP’s hybridization approach holds immense promise.

NIP architectures are continually evolving, incorporating more advanced, higher-order relational features. As NIPs become more sophisticated, better capturing more diverse physcis, HackNIP’s performance may also improve. Meanwhile, such sophistication often incurs significant computational cost. One pragmatic alternative may be to extract features from multiple NIPs, each optimized for different orders of relational complexity, and then apply downstream, feature-based ML models. Analogous to emerging multi-agent strategies in the large-language model community, multi-NIP featurization strategies could offer a cost-effective alternative to building a single, computationally intensive NIP. Moreover, as in the hybridization approach adopted by HackNIP---where separate ML models divide roles (one model converts raw structures into physics-rich representations while the other maps those representations to properties)---combining several specialized deep NNs with shallow ML models may further help balance predictive performance and computational efficiency. 

Overall, we demonstrate that assigning distinct roles to $\geq$2 ML models can outperform a single, specialized end-to-end model on materials-property prediction tasks. Specifically, the deep-shallow hybrid framework, HackNIP (ORB + MODNet), matches or exceeds state-of-the-art performance on the Matbench leaderboard, often achieving accuracy within the inherent uncertainty of DFT data relative to experiment. It is lightweight enough to run on free-tier cloud computing services for datasets up to 10$^5$ samples, and data-efficient enough to outperform direct fine-tuning on datasets up to 10$^4$ samples---the size range within which most materials falls---property databases, while remaining broadly applicable across downstream tasks. These include molecular versus extended structures, computational versus experimental datasets, multi-phase phenomena, and both classification and regression problems. Our detailed analysis indicates that closer investigation into performance's dependence on embedding depths used for feature extraction has the potential to provide more understandings on structure-property relationships, offering guidelines for future model development. The balanced robustness of this hybridization strategy could offer a fast yet reliable solution for many materials-science ML applications and is expected to significantly lower the resource-barrier to leveraging advanced ML capabilities in materials design.

\section{Related works}
\label{sec:related_works}

\subsection{Representation learning for molecules}
There have been various efforts to obtain generalized molecular representations through large-scale pretraining. Initial research efforts included attempts to utilize self-supervised learning on molecular topologies. Contrastive learning is a self-supervised approach that trains models to distinguish between similar and dissimilar data pairs by pulling representations of similar inputs closer and pushing those of dissimilar ones apart. GraphCL~\cite{you2020GraphCL} proposed to learn molecular representations through graph-level contrastive learning. MolCLR \cite{wang2022MolCLR} expanded this idea on larger dataset of approximately 10 million molecules. GEM \cite{fang2022GEM} integrated 3D structural information from approximately 20 million molecules for better molecular representations. Furthermore, LACL \cite{park2023LACL} proposed a self-supervised approach to learn common node-level representations between conformations of different fidelity.

Inspired by the fact that molecules can be represented as SMILES~\cite{weininger1988smiles, weininger1989smiles} strings, there have been attempts to leverage large-language models (LLMs) to obtain generalized molecular representations. ChemBERTa-2~\cite{ahmad2022ChemBERTa-2}, based on the RoBERTa~\cite{liu2019RoBERTa} architecture, was trained on approximately 77 million molecular SMILES to enhance the understanding of SMILES itself. It also reveals that representations from LLMs such as GPT \cite{radford2018GPT1, radford2019GPT2} and LLaMA~\cite{touvron2023Llama, touvron2023llama2} without fine tuning are also effective in the fields of chemistry and materials science~\cite{sadeghi2024GPTvsLlama_MoleculeEmbedding}. However, due to the inherent limitations of information in SMILES strings, difficulties in directly learning 3D structural information of molecules have also been noted.

Despite these advancements, the aforementioned models are not able to be extended to achieve comprehensive representation for extended structures since they are trained only for molecules. Therefore, comprehensively representing and predicting the microscopic and macroscopic properties of complex solid materials—such as the periodicity of crystal structures, the disorder in amorphous materials, or various defects within materials—still presents a significant challenge.

\subsection{Representation learning for crystals}
 
Unlike molecules that can be effectively represented by topology information, crystals are heavily dependent on their 3D structures. Consequently, the cost of acquiring unlabeled data for crystals is substantial. This leads to much sparser datasets compared to molecular data, resulting in relatively fewer representation learning studies for crystals. For example, Magar et al. \cite{magar2022CrystalTwins} adapt a twin GNN and learn representations by implementing Barlow Twins~\cite{zbontar2021barlowtwins} and SimSiam~\cite{chen2020SimSiam} frameworks. Koker et al. \cite{koker2022CrystalCLR} introduced contrastive learning strategy to improve performance of GNNs for material properties prediction. Recently, with a significant increase in available simulation datasets, the reliance on self-supervised approaches has lessened compared to before. However, still demonstrate that using generative diffusion as a pre-training step before performing supervised learning can achieve stronger performance, confirming that strategies for learning useful representations from limited data remain effective.

\subsection{Machine learning interatomic potentials}
MLIPs represent a class of interatomic potentials constructed through the application of machine learning algorithms. These algorithms are trained to map the relationship between the atomic structure of a material or molecule and its corresponding potential energy. The fundamental objective of an MLIP is to accurately approximate the potential energy surface, which governs the interactions between atoms and dictates a wide range of material properties and behaviors. 

A landmark contribution that catalyzed a surge of interest in MLIPs was Behler-Parrinello Neural Network (BPNN) and Gaussian Approximation Potential (GAP) \cite{behler2007BehlerParrinelloNN, bartok2010GaussianApproximationPotentials}. A significant trend in modern MLIP development is the adoption of GNNs. In this paradigm, atomic structures are naturally represented as graphs, where atoms serve as nodes and interatomic bonds or interactions (within a certain cutoff) are represented as edges. GNNs' end-to-end learning capability results in more flexible, powerful, and generalizable models. The principled incorporation of physical symmetries into ML models has deepened with the development of equivariant neural network architectures \cite{satorras2021EGNN}. The explicit embedding of equivariance directly into the neural network architecture often leads to more data-efficient learning, as the model does not need to learn these fundamental symmetries from scratch from the data~\cite{batzner2022NequIP, batatia2022MACE, liao2023EquiformerV2}. It also typically results in better generalization to unseen structures and more accurate predictions of vectorial and tensorial properties. With the recent emergence of large datasets \cite{barroso-luque2024OpenMaterials2024, deng2023CHGNet} that can effectively train large-scale MLIPs, they have successfully bridged the long-standing gap between the accuracy of quantum mechanics and the efficiency of classical potentials, enabling studies at unprecedented scales of system size, timescale, and structural or chemical complexity, all while maintaining a level of accuracy that approaches that of the underlying first-principles methods \cite{dembitskiy2025LiionMigrationMLIPNEB}.

\subsection{Transfer learning}
Transfer learning is a machine learning method where knowledge from a pretrained model is applied to a new, related problem. Instead of training a new model from scratch, this method leverages the features or weights learned by the existing model, either by fine-tuning them or using them directly for a new task, thereby reducing training time and enabling high performance even with less data. For example, a cross-property transfer learning was proposed to address data scarcity challenge in materials science applications \cite{gupta2021CrosspropertyDeepTransferLearning}. As NIP development has matured and many developers now provide open access to hidden-layer embeddings, these representations are widely available for transfer learning, and recent work using universal NIPs such as M3GNet~\cite{chen2022M3GNet} and MACE~\cite{batatia2022MACE} as descriptors has achieved predictive performance on par with state-of-the-art NMR chemical-shift prediction methods~\cite{shiota2024UniversalNeuralNetworkasDescriptors}. A power-law scaling for simulation-to-real transfer learning in materials science was demonstrated to guide strategic materials database development and resource allocation \cite{minami2025ScalingLawSim2RealTransferLearning}.









\section{Methods}
\subsection{Datasets}
We leveraged a diverse suite of open‐access materials and molecular benchmarks. The datasets included: Matbench (comprising exfoliation energy; the frequency of the last phonon peak in the phonon density of states; refractive index; shear and bulk moduli; perovskite formation energy; band gap; and formation energy), MoleculeNet \cite{wu2018MoleculeNet} (beta-secretase 1 inhibition \cite{subramanian2026BACEdataset}, blood-brain barrier penetration \cite{martins2012BBBPdataset}, and clinical toxicity \cite{gayvert2016ClinToxdataset}), a curated compilation of superconducting critical temperatures for compounds with available structural data (3DSC$_\mathrm{MP}$)~\cite{sommer20233DSCDatasetSuperconductors}, Li diffusivity in amorphous solids calculated via \textit{ab initio} molecular dynamics simulation \cite{zheng2024abinitioDiffusivity}, and Coulombic efficiency data for electrolyte precursor combinations \cite{kim2023ElectrolyteDesign}. 

\subsection{Neural network interatomic potentials and conventional machine learning models}
We used NIPs to extract hierarchical, layer-wise embeddings from each atomic structure and then applied conventional ML models to map these embeddings to their corresponding properties. We evaluated five hybrid deep–shallow frameworks constructed from different combinations of NIPs and conventional ML models. The pretrained NIP foundation models tested were ORB (“orb-v2”)~\cite{neumann2024ORBv2}, EquiformerV2 (“eqV2 M”)~\cite{liao2023EquiformerV2}, and MACE (“MACE-MP-0a medium”)~\cite{batatia2022MACE}, and the shallow ML models evaluated were MODNet~\cite{debreuck2021MODNet}, XGBoost, and MLP. The five combinations were ORB–MODNet, ORB–XGBoost, ORB–MLP, EquiformerV2–MODNet, and MACE–MODNet. Under a chosen set of hyperparameters (detailed below), ORB–MODNet achieved the best overall performance on Matbench tasks and was selected as the model HackNIP architecture for this study; note that other combinations might outperform it under different settings.

\subsection{Data preprocessing}
Raw inputs were first sorted into two categories: isolated molecules and extended structures (i.e., atomic assemblies that go beyond discrete molecules, including both crystalline lattices and amorphous solids). Molecular structures were generated from SMILES strings and subsequently relaxed via geometry optimization using ORB prior to feature extraction. Extended structures were expanded into supercells with each lattice vector exceeding 10 \AA\ for ORB~\cite{neumann2024ORBv2} and EquiformerV2~\cite{liao2023EquiformerV2}, and 12 \AA\ for MACE~\cite{batatia2022MACE} to ensure complete neighbor connectivity within the largest model cutoff, considering that edges are created using distance thresholds of 10 \AA, 6 \AA, and 12 \AA for chosen versions of ORB, EquiformerV2 and MACE foundational models, respectively, during graph construction. No further relaxation was performed, as all extended‐structure inputs were sourced from DFT‐relaxed databases.

\subsection{Feature extraction}
Each atomic structure was passed through NIP foundation models truncated at a specified network depth to yield layer-wise embeddings. In all cases, the raw atomic coordinates and species were converted into the model’s internal representation, propagated through the shared encoder, and then through the first \textit{L} layers of the neural backbone (where \textit{L} was varied between experiments). The resulting atom-level features were finally averaged over all atoms to produce a fixed-length descriptor vector. By extracting these mean-pooled embeddings at different depths (1 $\leq$ \textit{L} $\leq$ 15 for ORB, 0 $\leq$ \textit{L} $\leq$ 9 for EquiformerV2, and 1 $\leq$ \textit{L} $\leq$ 2 for MACE), we obtained a hierarchical set of features that capture progressively more global structural information for downstream analysis. The feature dimensionality was 256 for every ORB layer, 128 for every EquiformerV2 layer, and 128- and 256-dimensional for the first and second MACE layers, respectively.

\subsection{Feature-based machine learning}
For downstream property prediction, we treated layer-wise embeddings from each NIP layer as a fixed feature vector and trained three types of ML models: MODNet, XGBoost, and MLP. MODNet models were constructed with four layers with 256 neurons each. XGBoost models were fit with 256 trees, a maximum depth of 4, a learning rate of 0.1, and 80\% subsampling of both rows and columns. MLPs employed two hidden layers of 100 units each and were trained for up to 200 epochs. For each model and each embedding depth \(L\), we performed five-fold experiments (using Matbench’s standard seed of 18012019), except for the Coulombic-efficiency task with a limited dataset size, for which we used five independent 70/30 splits to match prior studies. We identified the best-performing embedding depth \(L^*\) and considered its corresponding predictive performance to represent the HackNIP architecture's performance. 

\subsection{Hyperparameter optimization}

For the model HackNIP architecture (ORB + MODNet), we performed a focused Optuna \cite{akiba2019optuna} search over MODNet hyperparameters---batch size, learning rate, number of input features, network depth, and layer width---running 50 trials at each of five ORB depths around the best-performing embedding depths (from \(L^*-2\) to \(L^*+2\)). Each trial was scored by the MAE averaged across outer folds, with early stopping after 10 non-improving trials. The best hyperparameters were used to train final MODNet models for evaluation. Five-fold experiments were performed to calculate the mean and standard deviation of the MAEs. Parity plots were generated for the fold whose MAE was closest to the average MAE across all tested folds.

\subsection{Fine-tuning}

For property-specific fine-tuning---used to compare against the HackNIP architecture with ORB feature extraction---we employed the same ORB foundation model (i.e., orb-v2) initialized with pretrained weights and appended a single-output MLP head. The MLP head consisted of two 128-unit layers (256 $\rightarrow$ 128 $\rightarrow$ 128) with SiLU activations, followed by a final 128→1 linear layer; all linear weights use Xavier uniform initialization and biases were zero. In standard fine-tuning runs, the entire dataset (100\%) was used; for data-efficiency experiments, we sampled reproducible random fractions of the data to assess performance as a function of training set size. All ORB encoder layers were unfrozen except for the initial edge-featurizer and final decoder blocks, which remained fixed; only the MLP head and the unfrozen backbone parameters were optimized. Training proceeded for 200 epochs with AdamW using discriminative learning rates (1 $\times$ 10$^{-5}$ for the backbone and 1 $\times$ 10$^{-3}$ for the head) and a weight decay of 1 $\times$ 10$^{-4}$. The model achieving the lowest validation MAE was checkpointed and then evaluated on the held-out test split to report final test MAE. All model-training experiments were conducted on a workstation equipped with ten NVIDIA RTX 6000 Ada graphics cards and dual Intel Ice Lake–class processors.

\section*{Conflict of interest}
The authors declare that there are no competing interests. 
\section*{Data availability}
The data that support the findings of this study are available at Matbench~\cite{dunn2020MatBench}, MoleculeNet~\cite{wu2018MoleculeNet}, 3DSC~\cite{sommer20233DSCDatasetSuperconductors}, MPContribs~\cite{zheng2024abinitioDiffusivity} and Supplementary information of \cite{kim2023ElectrolyteDesign}.
\section*{Code availability}
The Python codes used for this work are available on GitHub at \url{https://github.com/parkyjmit/HackNIP}.

\section*{Author contribution}
S.Y.K., Y.J.P., and J.L. designed the study. S.Y.K. carried out key investigations and performed the analyses alongside Y.J.P. and J.L. S.Y.K. and Y.J.P. drafted the original manuscript. Y.J.P. curated the data. J.L. supervised the project. All authors reviewed and edited the manuscript, approved the final version, and agree to be accountable for all aspects of the work.

\section*{Acknowledgments}
We acknowledge support from Samsung SDI. S.Y.K. gratefully acknowledges support from the Kwanjeong Fellowship. 


\bibliography{references}

\newpage
\bmhead{Supplementary information}
\renewcommand{\thefigure}{S\arabic{figure}}
\renewcommand{\thetable}{S\arabic{table}}
\setcounter{figure}{0}
\setcounter{table}{0}
\setcounter{section}{0}


\section{Exploration of model performance on Matbench tasks}\label{secA1}

\begin{figure*}[h!]
    \centering
    \includegraphics[width=\textwidth]{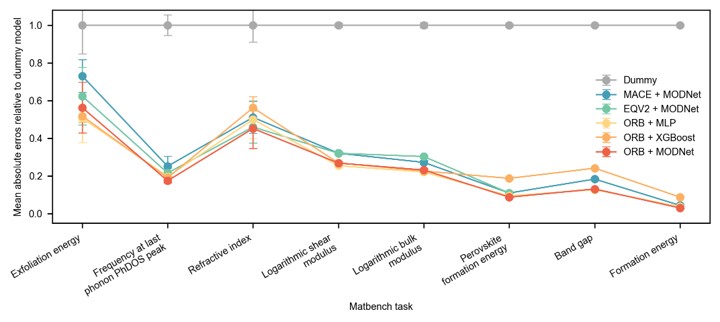}
    \caption{\textbf{HackNIP combinations and performance across eight Matbench tasks.} HackNIPs built from various NIP foundation models (ORB, EquiformerV2, and MACE) and shallow regressors (MODNet, XGBoost, MLP) are shown, with relative mean absolute errors against a dummy model that predicts the training‐set mean. Mean absolute errors are averaged over five-fold experiments. When the shallow regressor is fixed to MODNet, ORB outperforms EquiformerV2 and MACE. When the NIP model is fixed to ORB, MODNet and MLP outperform XGBoost, and MODNet training is more cost-efficient than MLP training, for the chosen hyperparameters. In other settings, different combinations may outperform.}
    \label{fig:NIP-ml_combination}
\end{figure*}

\begin{table*}[h!]
    \centering
    \caption{\textbf{Summary of optimized model parameters for Matbench tasks.} ORB embedding depths used for feature extraction and optimized MODNet hyperparameters are shown.}
    \includegraphics[width=\textwidth]{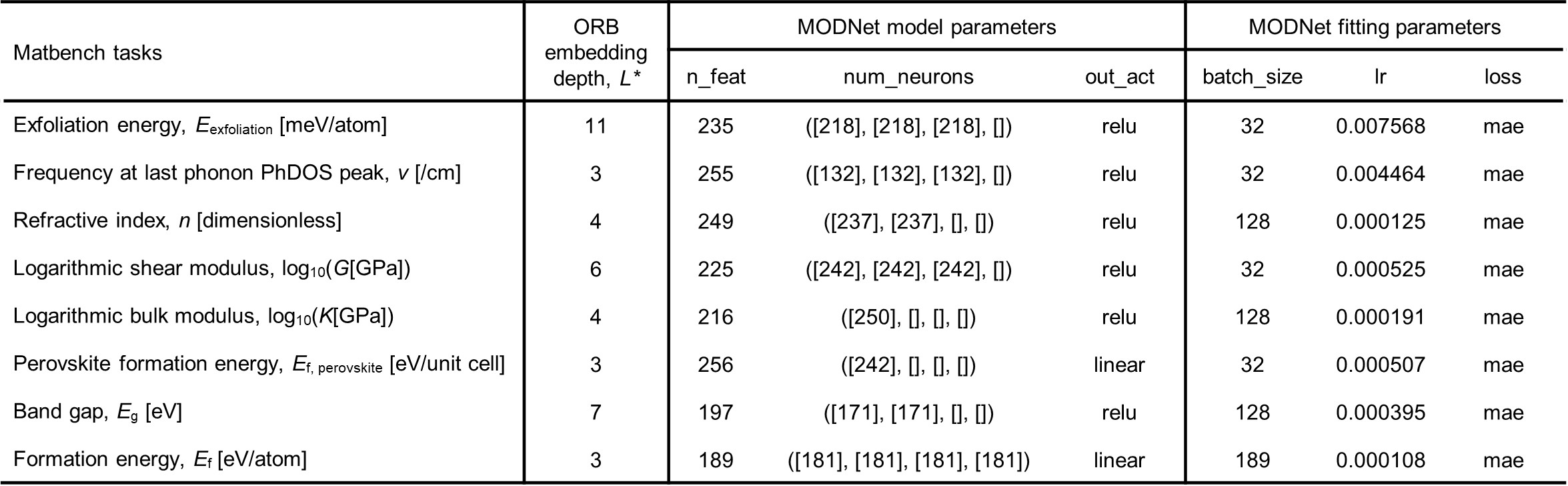}
    \label{tab:s_matbench_opted-hp}
\end{table*}

\begin{figure*}[h!]
    \centering
    \includegraphics[width=\textwidth]{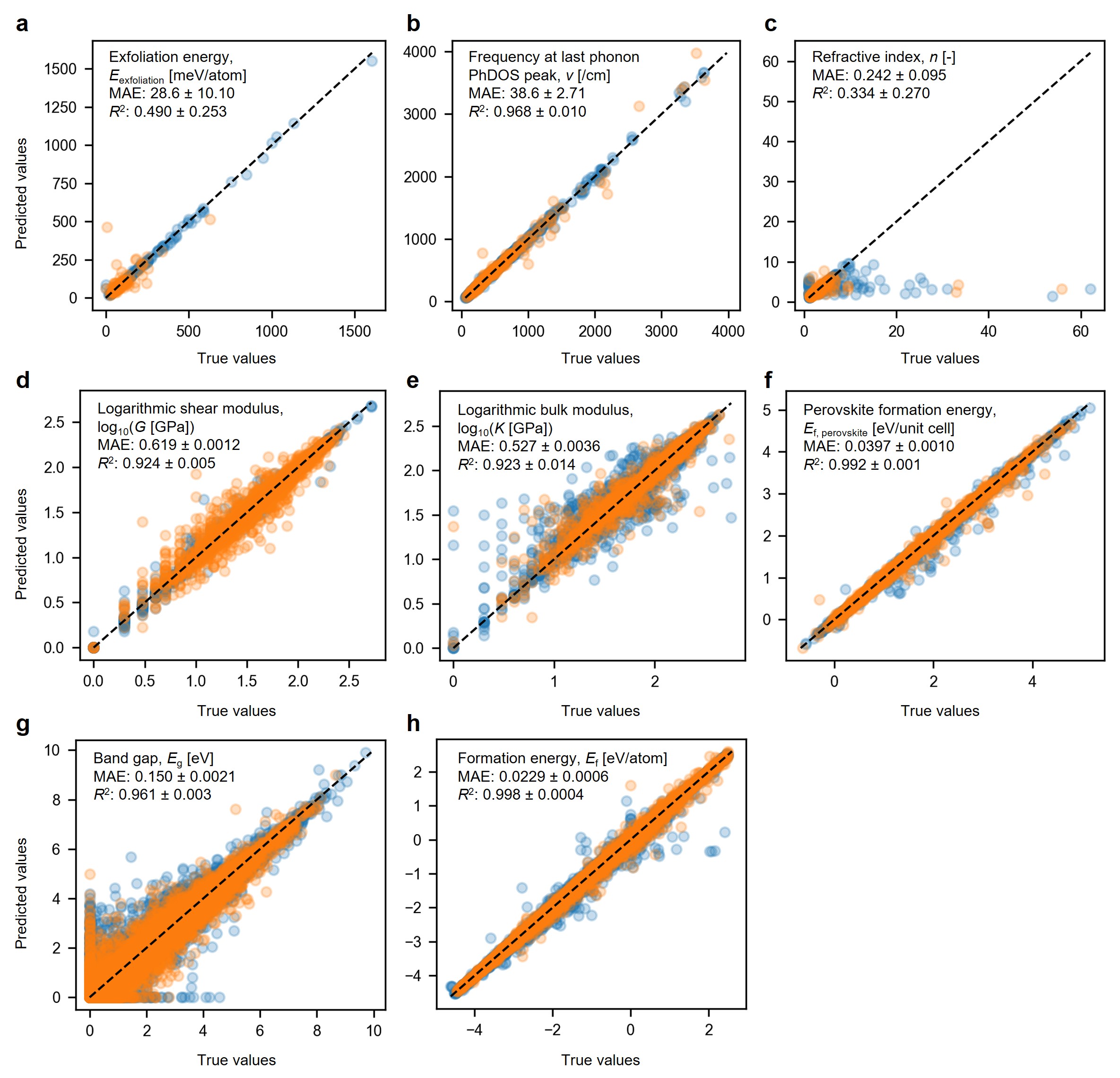}
    \caption{\textbf{Parity plots for eight Matbench tasks.} \textbf{a,} Exfoliation energy. \textbf{b,} Frequency at the last phonon peak in the phonon density of states (PhDOS). \textbf{c,} Refractive index. \textbf{d,} Logarithmic shear modulus. \textbf{e,} Logarithmic bulk modulus. \textbf{f,} Perovskite formation energy. \textbf{g,} Band gap. \textbf{h,} Formation energy. Mean absolute errors (MAEs) and coefficients of determination ($R^2$) are provided. Blue: Train; Orange: Test.}
    \label{fig:s_matbench_parity}
\end{figure*}

\mbox{~}
\clearpage
\newpage
\section{Exploration of model performance on diverse downstream tasks}\label{secA2}

\begin{table*}[h!]
    \centering
    \caption{\textbf{Summary of optimized model parameters for diverse downstream tasks.} ORB embedding depths used for feature extraction and optimized MODNet hyperparameters are provided.}
    \includegraphics[width=\textwidth]{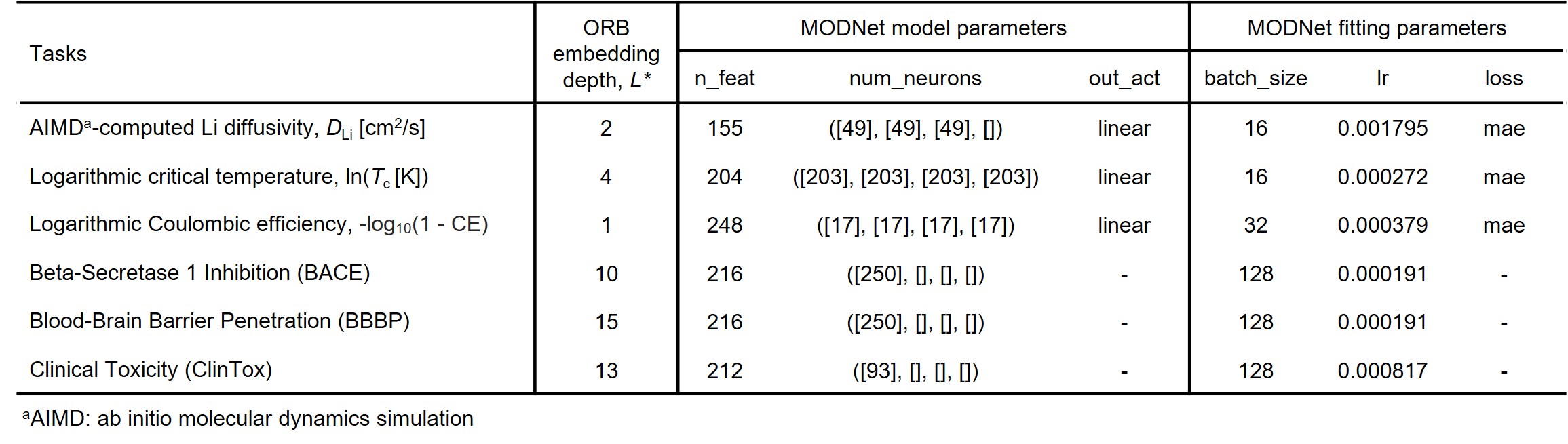}
    \label{tab:s_general_opted-hp}
\end{table*}

\mbox{~}
\clearpage
\newpage

\section{Embedding depth for feature extraction and model performance}\label{secA3}

\begin{figure*}[h!]
    \centering
    \includegraphics[width=\textwidth]{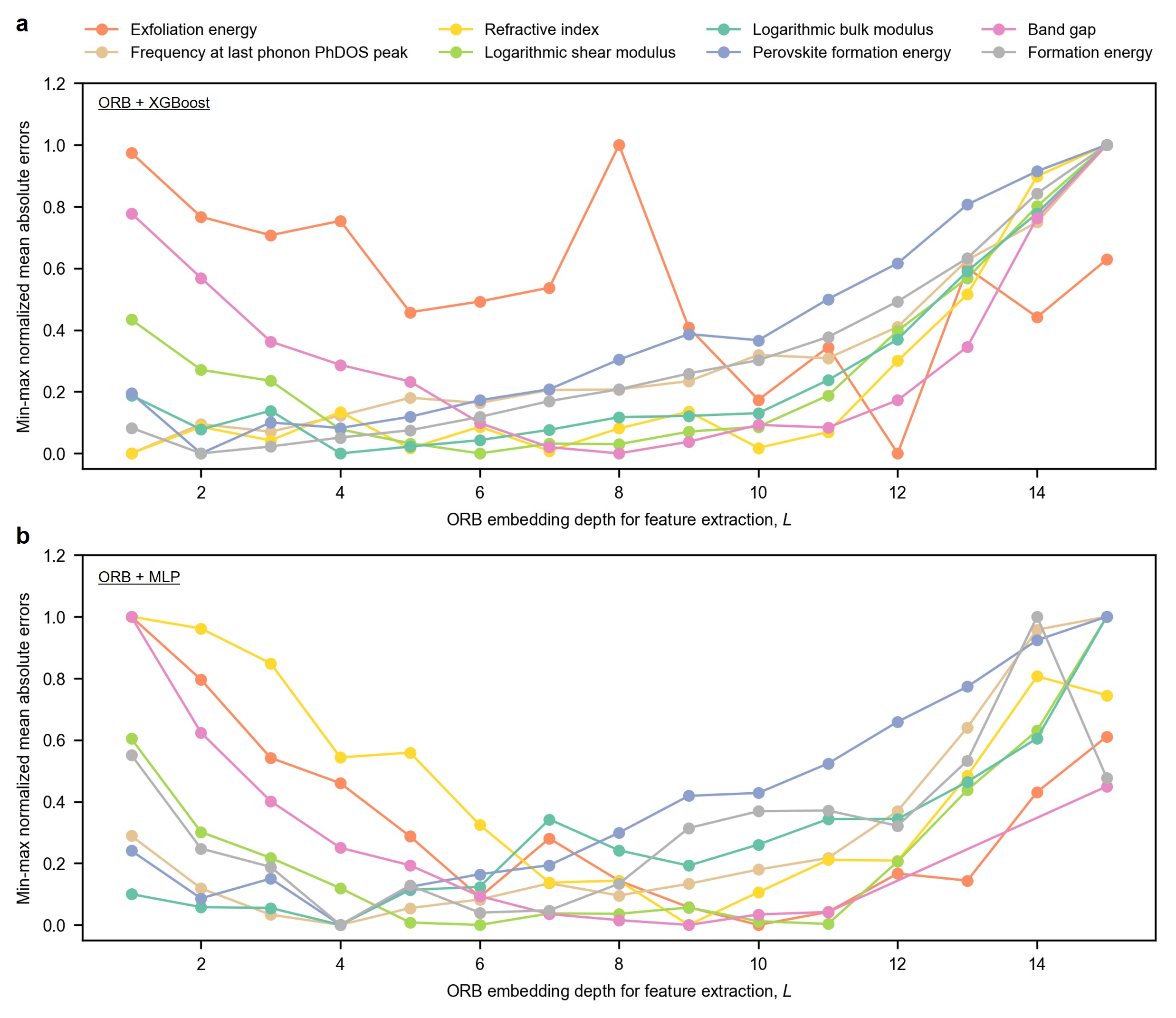}
    \caption{\textbf{HackNIP performance as a function of ORB embedding depth used for feature extraction. }Two shallow regressors are shown: \textbf{a,} extreme gradient boosting algorithm (XGBoost) and \textbf{b,} multilayer perceptron (MLP). Mean absolute errors are min–max normalized across embedding depths for each Matbench task. Five-fold experiments were performed for each point and only the mean values are presented for clarity.}
    \label{fig:s_layer_xgb-mlp}
\end{figure*}






\end{document}